\documentclass[pre,twocolumn,aps,showpacs]{revtex4}
\newcommand{\bec}[1]{\mbox{\boldmath $ #1$}}

\usepackage{graphicx}
\usepackage{verbatim}

\begin{document}
\title{Large-scale instability in a sheared nonhelical turbulence:
formation of vortical structures}
\author{Tov Elperin}
\email{elperin@bgu.ac.il}
\homepage{http://www.bgu.ac.il/~elperin}
\author{Ilia Golubev}
\email{golubev@bgu.ac.il}
\author{Nathan Kleeorin}
\email{nat@menix.bgu.ac.il}
\author{Igor Rogachevskii}
\email{gary@bgu.ac.il}
\homepage{http://www.bgu.ac.il/~gary}
\affiliation{The Pearlstone Center for Aeronautical Engineering
Studies, Department of Mechanical Engineering, Ben-Gurion University
of the Negev, Beer-Sheva 84105, P. O. Box 653, Israel}
\date{\today}
\begin{abstract}
We study a large-scale instability in a sheared nonhelical
turbulence that causes generation of large-scale vorticity. Three
types of the background large-scale flows are considered, i.e., the
Couette and Poiseuille flows in a small-scale homogeneous
turbulence, and the "log-linear" velocity shear in an inhomogeneous
turbulence. It is known that laminar plane Couette flow and
antisymmetric mode of laminar plane Poiseuille flow are stable with
respect to small perturbations for any Reynolds numbers. We
demonstrate that in a small-scale turbulence under certain
conditions the large-scale Couette and Poiseuille flows are unstable
due to the large-scale instability. This instability causes
formation of large-scale vortical structures stretched along the
mean sheared velocity. The growth rate of the large-scale
instability for the "log-linear" velocity shear is much larger than
that for the Couette and Poiseuille background flows. We have found
a turbulent analogue of the Tollmien-Schlichting waves in a
small-scale sheared turbulence. A mechanism of excitation of
turbulent Tollmien-Schlichting waves is associated with a combined
effect of the turbulent Reynolds stress-induced generation of
perturbations of the mean vorticity and the background sheared
motions. These waves can be excited even in a plane Couette flow
imposed on a small-scale turbulence when perturbations of mean
velocity depend on three spatial coordinates. The energy of these
waves is supplied by the small-scale sheared turbulence.
\end{abstract}


\maketitle

\section{Introduction}

Large-scale vortical structures are universal features observed in
geophysical, astrophysical and laboratory flows (see, e.g.,
\cite{L83,P87,C94,GLM97,T98,RAO98}). Formation of vortical
structures is related to the Prandtl secondary flows (see, e.g.,
\cite{P52,T56,P70,B87}). A lateral stretching (or ''skewing") by an
existing shear generates streamwise vorticity that results in
formation of the first kind of the Prandtl secondary flows. In
turbulent flow the large-scale vorticity is generated by the
divergence of the Reynolds stresses. This mechanism determines the
second kind of the Prandtl turbulent secondary flows \cite{B87}.

The generation of large-scale vorticity in a homogeneous nonhelical
turbulence with an imposed large-scale linear velocity shear has
been recently studied in \cite{EKR03}. Let us discuss a mechanism of
this phenomenon. The equation for the mean vorticity ${\bf W} =
\bec{\nabla} {\bf \times} {\bf U}$ read
\begin{eqnarray}
{\partial {\bf W} \over \partial t} = \bec{\nabla} {\bf \times}
({\bf U} {\bf \times} {\bf W} + {\bf F} - \nu \bec{\nabla} {\bf
\times} {\bf W})  \;,
\label{W10}
\end{eqnarray}
where ${\bf U}$ is the mean fluid velocity, ${\bf F}_i = - \nabla_j
\, \langle u_i u_j \rangle$ is the effective force caused by
velocity fluctuations, ${\bf u}$, and $ \nu$ is the kinematic
viscosity. The first term, ${\bf U} {\bf \times} {\bf W}$, in
Eq.~(\ref{W10}) determines laminar effects of the mean vorticity
production caused by the sheared motions, while the effective force
${\bf F}$ determines the turbulent effects on the mean fluid flow.
Let us consider a simple large-scale linear velocity shear ${\bf
U}^{(s)} = (0, Sx, 0)$ imposed on the small-scale nonhelical
turbulence. The equation for the perturbations of the mean
vorticity, $\tilde{\bf W} = (\tilde{W}_x(z), \tilde{W}_y(z), 0)$,
reads
\begin{eqnarray}
{\partial \tilde{W}_x \over \partial t} &=& S \, \tilde{W}_y +
\nu_{_{T}} \tilde{W}''_x \;,
\label{E2}\\
{\partial \tilde{W}_y \over \partial t} &=& - \beta_0 \, S \, l_0^2
\, \tilde{W}''_x + \nu_{_{T}} \tilde{W}''_y  \;,
\label{E3}
\end{eqnarray}
(see \cite{EKR03}), where $\tilde{W}'' = \partial^2 \tilde{W}
/\partial z^2$, $\, \nu_{_{T}}$ is the turbulent viscosity, $l_0$ is
the maximum scale of turbulent motions and the parameter $\beta_0$
is of the order of 1, and depends on the scaling exponent of the
correlation time of the turbulent velocity field (see Sect. II). A
solution of Eqs.~(\ref{E2}) and~(\ref{E3}) has the form $ \propto
\exp(\gamma t + i K_z z)$, where the growth rate of the large-scale
instability is given by $\gamma = \sqrt{\beta_0} \, S \, l_0 \, K_z
- \nu_{_{T}} \, K_z^2$ and $K_z$ is the wave number. The maximum
growth rate of perturbations of the mean vorticity, $ \gamma_{\rm
max} = \beta_0 \, (S \, l_0)^2 / 4 \nu_{_{T}}$, is attained at $ K_z
= K_m = \sqrt{\beta_0} \, S \, l_0 /2 \nu_{_{T}}$. This corresponds
to the ratio $\tilde{W}_y / \tilde{W}_x = \sqrt{\beta_0} \, l_0 \,
K_m \approx S \, \tau_0$, where the time $\tau_0 = l_{0} / u_0$ and
$u_0$ is the characteristic turbulent velocity in the maximum scale
$l_{0}$ of turbulent motions. Note that in a laminar flow this
instability does not occur.

The mechanism of this instability is as follows (see \cite{EKR03}
for details). The first term, $S \tilde{W}_y = ({\bf
W}^{(s)}\cdot\bec{\nabla})~\tilde{U}_x$, in Eq.~(\ref{E2})
determines a ''skew-induced" generation of perturbations of the mean
vorticity $\tilde{W}_x$ by stretching of the equilibrium mean
vorticity ${\bf W}^{(s)}= (0,0,S)$, where $\tilde{\bf U}$ are the
perturbations of the mean velocity. In particular, the mean
vorticity $\tilde{W}_x {\bf e}_x$ is generated from $\tilde{W}_y
{\bf e}_y$ by equilibrium shear motions with the mean vorticity
${\bf W}^{(s)}$, whereby $\tilde{W}_x {\bf e}_x \propto ({\bf
W}^{(s)} \cdot \bec{\nabla}) \tilde{U}_x {\bf e}_x \propto
\tilde{W}_y {\bf e}_y \times {\bf W}^{(s)} $. Here ${\bf e}_x$,
${\bf e}_y$ and ${\bf e}_z$ are the unit vectors along $x$, $y$ and
$z$ axes, respectively. On the other hand, the first term, $-
\beta_0 \, S \, l_0^2 \, \tilde{W}''_x$, in Eq.~(\ref{E3})
determines a ''Reynolds stress-induced" generation of perturbations
of the mean vorticity $\tilde{W}_y$ by the Reynolds stresses.  In
particular, this term is determined by $ (\bec{\nabla} {\bf \times}
{\bf F})_y$. This implies that the component of the mean vorticity
$\tilde{W}_y {\bf e}_y $ is generated by an effective anisotropic
viscous term $ \propto - l_0^2 \, \Delta \, (\tilde{W}_x {\bf e}_x
\cdot \bec{\nabla}) \, {U}^{(s)}(x) {\bf e}_y \propto - l_0^2 \, S
\, \tilde{W}''_x {\bf e}_y .$ This instability is caused by a
combined effect of the sheared motions (''skew-induced" generation)
and the ''Reynolds stress-induced" generation of perturbations of
the mean vorticity.

The mechanism for this large-scale instability in a sheared
nonhelical homogeneous turbulence is different from that discussed
in \cite{MST83,KMT91,CMP94}, where the generation of large-scale
vorticity in the helical turbulence occurs due to hydrodynamic alpha
effect. The latter effect is associated with the hydrodynamic
helicity of turbulent flow. In a nonhelical homogeneous turbulence
this effect does not occur.

The large-scale instability in a nonhelical homogeneous turbulence
has been studied in \cite{EKR03} only for a simple case of unbounded
turbulence with an imposed linear velocity shear and when the
perturbations of the mean vorticity depend on one spatial variable
$z$. In this study the theoretical approach proposed in \cite{EKR03}
is further developed and applied for comprehensive investigation of
the large-scale instability for different situations with nonuniform
shear, inhomogeneous turbulence and a more general form of the
perturbations of the mean vorticity $\tilde{\bf W}({\bf r})$ that
depends on three spatial variables.

In the present study we consider three types of the background
large-scale flows, i.e., the Couette flow (linear velocity shear)
and Poiseuille flow (quadratic velocity shear) in a small-scale
homogeneous turbulence, and the "log-linear" velocity shear in an
inhomogeneous turbulence. We have derived new mean-field equations
for perturbations of large-scale velocity which depend on three
spatial coordinates in a small-scale sheared turbulence, for a
nonuniform background large-scale velocity shear and for an
arbitrary scaling of the correlation time $\tau(k)$ of the turbulent
velocity field.

The stability of the laminar Couette and Poiseuille flows in a
problem of transition to turbulence has been studied in a number of
publications (see, e.g., \cite{DR81,SH01,CJJ03,BOH88,REM03,ESH07},
and references therein). It is known that laminar plane Couette flow
and antisymmetric mode of laminar plane Poiseuille flow are stable
with respect to small perturbations for any Reynolds numbers. A
symmetric mode of laminar plane Poiseuille flow is stable when the
Reynolds number is less than 5772 \cite{CJJ03}. In laminar flows the
Tollmien-Schlichting waves can be excited. The molecular viscosity
plays a destabilizing role in laminar flows which promotes the
excitation of the Tollmien-Schlichting waves (see, e.g.,
\cite{SH01}). These waves are growing solutions of the
Orr-Sommerfeld equation.

In the present study we have found a turbulent analogue of the
Tollmien-Schlichting waves. These waves are excited by a small-scale
sheared turbulence, i.e., by a combined effect of the turbulent
Reynolds stress-induced generation of perturbations of the mean
vorticity and the background sheared motions. The energy of these
waves is supplied by the small-scale sheared turbulence. We
demonstrate that the off-diagonal terms in the turbulent viscosity
tensor play a crucial role in the excitation of the turbulent
Tollmien-Schlichting waves. These waves can be excited even in a
plane Couette flow imposed on a small-scale turbulence when
perturbations of velocity depend on three spatial coordinates. When
perturbations of large-scale velocity depend on one or two spatial
coordinates the turbulent Tollmien-Schlichting waves can not be
excited in a sheared turbulence. In the present study we show that
the large-scale Couette and Poiseuille flows imposed on a
small-scale turbulence can be unstable with respect to small
perturbations. The critical effective Reynolds number (based on
turbulent viscosity) required for the excitation of this large-scale
instability, is of the order of 200.

This paper is organized as follows. In Sect. II the governing
equations are formulated. In Sect. III we consider a homogeneous
turbulence with a large-scale linear velocity shear (Couette flow),
while in Sect. IV we study a homogeneous turbulence with a
large-scale quadratic velocity shear (Poiseuille flow). In Sect. V
we investigate formation of large-scale vortical structures in an
inhomogeneous turbulence with an imposed nonuniform velocity shear.
Finally, we draw conclusions in Sec.~VI.

\section{Governing equations}

The equation for the mean velocity ${\bf U}$ in incompressible flow
reads
\begin{eqnarray}
\bigg( {\partial \over \partial t} + {\bf U} \cdot \nabla \bigg)U_i
= - {\nabla_i P \over \rho}  + \nabla_j \, \langle u_i u_j \rangle +
\nu \Delta U_i \;, \label{B2}
\end{eqnarray}
where ${\bf U}$ is the mean velocity, $P$ is the mean pressure and
$\nu$ is the kinematic viscosity. The effect of turbulence on the
mean flow is determined by the Reynolds stresses $\langle u_i u_j
\rangle$, where ${\bf u}$ are the fluid velocity fluctuations.

We consider a turbulent flow with an imposed mean velocity shear
$\nabla_i {\bf U}^{(s)}$, where ${\bf U}^{(s)}$. In order to study a
stability of this equilibrium we consider perturbations $\tilde{\bf
U}$ of the mean velocity, i.e., the total mean velocity is ${\bf U}
= {\bf U}^{(s)} + \tilde{\bf U}$. Thus, the linearized equation for
the small perturbations of the mean velocity is given by
\begin{eqnarray}
\bigg({\partial \over \partial t} + {\bf U}^{(s)} \cdot
\bec{\nabla}\bigg) \tilde{U}_i &+& (\tilde{\bf U} \cdot
\bec{\nabla})U^{(s)}_i = -{\nabla_i \tilde{P} \over \rho}  + F_i
\nonumber\\
&+& \nu \Delta \tilde{U}_i \;, \label{B4}
\end{eqnarray}
where $F_i = - \nabla_j \, f_{ij}(\tilde{\bf U})$ is the effective
force, $f_{ij} = \langle u_i u_j \rangle$ and $\tilde{P}$ are the
perturbations of the fluid pressure. Equation~(\ref{B4}) is derived
by subtracting Eq.~(\ref{B2}) written for the equilibrium velocity
${\bf U}^{(s)}$ from Eq.~(\ref{B2}) for the mean velocity ${\bf U}$.
We consider a simple large-scale velocity shear, so that ${\bf
U}^{(s)}$ is directed along $y$ direction and is non-uniform in $x$
direction, i.e., $ {\bf U}^{(s)} = (0, U^{(s)}_y(x), 0)$.

In order to obtain a closed system of equations, an equation for the
effective force $F_i = - {\nabla}_j f_{ij}(\tilde{\bf U})$ has been
derived in \cite{EKR03}, where
\begin{eqnarray}
f_{ij}(\tilde{\bf U}) &=& - 2 \nu_{_{T}} \, (\partial \tilde U)_{ij}
- l_0^2 \, \big[4 C_1 \, M_{ij} + C_2 \, (N_{ij} + H_{ij})
\nonumber\\
& & + C_3 \, G_{ij}\big] \;,
\label{B15}
\end{eqnarray}
$(\partial \tilde U)_{ij} = (\nabla_i \tilde U_{j} + \nabla_j \tilde
U_{i}) / 2$ and $l_0$ is the maximum scale of turbulent motions. The
tensors $\, M_{ij},$ $\, N_{ij} ,$ $\, H_{ij}$ and $G_{ij}$, in the
expression for the Reynolds stresses~(\ref{B15}) are given by:
\begin{eqnarray*}
M_{ij} &=& (\partial {U}^{(s)})_{im} ({\partial \tilde U})_{mj} +
(\partial {U}^{(s)})_{jm} ({\partial \tilde U})_{mi} \;,
\\
N_{ij} &=& \tilde{W}_n [\varepsilon_{nim} (\partial {U}^{(s)})_{mj}
+ \varepsilon_{njm} (\partial {U}^{(s)})_{mi}] \;,
\\
H_{ij} &=& {W}^{(s)}_n [\varepsilon_{nim} (\partial \tilde U)_{mj} +
\varepsilon_{njm} (\partial \tilde U)_{mi}]  \;,
\\
G_{ij} &=& {W}^{(s)}_i \tilde{W}_j + {W}^{(s)}_j \tilde{W}_i \;,
\end{eqnarray*}
$\varepsilon_{ijk}$ is the fully antisymmetric Levi-Civita tensor,
$(\partial {U}^{(s)})_{ij} = (\nabla_i {U}^{(s)}_{j} + \nabla_j
{U}^{(s)}_{i}) / 2 $ and the parameters $C_k$ in Eq.~(\ref{B15}) are
given below.

The effective force $F_i$ depends on the correlation time of the
turbulent velocity field $\tau(k)$, where $k$ is the wave number. In
the present study we derive a more general form of the effective
force $F_i$ for an arbitrary scaling of the correlation time
$\tau(k) = C \, \tau_0 \, (k / k_{0})^{-\mu}$ of the turbulent
velocity field, where $k_{0} = 1 / l_{0}$. To this end we use
Eq.~(20) derived in \cite{EKR03}. The value of the coefficient
$C=(q-1+\mu)/(q-1)$ corresponds to the standard form of the
turbulent viscosity in the isotropic turbulence, i.e., $\nu_{_{T}} =
\int \tau(k) \, [\langle {\bf u}^2 \rangle \, E(k)] \, dk = \tau_0
\, \langle {\bf u}^2 \rangle /3$. Here $E(k) = (q-1) \, k_{0}^{-1}
\, (k / k_{0})^{-q}$ is the energy spectrum of turbulence. For the
Kolmogorov's type background turbulence (i.e., for the turbulence
with a constant energy flux over the spectrum), the exponent
$\mu=q-1$ and the coefficient $C=2$. This case has been studied in
\cite{EKR03}. For a turbulence with a scale-independent correlation
time, the exponent $\mu=0$ and the coefficient $C=1$. The parameters
$C_k$ entering in the Reynolds stresses~(\ref{B15}) are given by
$C_1 = 2 C^2 \, (\mu^2 - 11 \, \mu + 28) / 315$, $ \, C_2 = - C^2 \,
(7 \, \mu + 1) / 90$ and $C_3 = - C^2 \, (\mu + 3) / 90$.

For the derivation of the effective force $F_i$ we use a procedure
outlined below (see \cite{EKR03} for details). Using the equation
for fluctuations of velocity written in a Fourier space, we derive
equation for the two-point second-order correlation function of the
velocity fluctuations $\langle u_i \, u_j\rangle$. We introduce a
background turbulence with zero gradients of the mean fluid
velocity. This background turbulence is determined by a stirring
force that is independent of gradients of the mean velocity. In this
study we use a model of isotropic, homogeneous and nonhelical
background turbulence. Then we subtract the equation for the
two-point second-order correlation function of the velocity
fluctuations $\langle u_i \, u_j\rangle^{(0)}$ written for the
background turbulence from the equation for $\langle u_i \,
u_j\rangle$. This yields the equation for the deviations from the
background turbulence.

The obtained second-moment equation include the first-order spatial
differential operators $\hat{\cal N}$  applied to the third-order
moments $M^{(III)}$. A problem arises how to close the equation,
i.e., how to express the third-order terms $\hat{\cal N} M^{(III)}$
through the lower moments $M^{(II)}$ (see, e.g.,
\cite{O70,MY75,Mc90}). To this end we use a spectral $\tau$
approximation which postulates that the deviations of the
third-moment terms from the contributions to these terms afforded by
the background turbulence are expressed through the similar
deviations of the second moments (see, e.g.,
\cite{O70,PFL76,KRR90,EKRZ02,EKR03}). A justification of the $\tau$
approximation for different situations has been performed in
numerical simulations and analytical studies in
\cite{BF02,FB02,BK04,BSM05,SSB07}.

We assume that the characteristic time of variation of the second
moment of velocity fluctuations is substantially larger than the
correlation time for all turbulence scales. This allows us to obtain
a steady state solution of the second moment equation for the
deviations from the background turbulence. Integration in ${\bf k}$
space allows us to determine the Reynolds stresses in the form of
Eq.~(\ref{B15}). Note that this form of the Reynolds stresses in a
turbulent flow with a mean velocity shear can be obtained even by
simple symmetry reasoning (see \cite{EKR03} for details).

In the next Sections we use Eq.~(\ref{B4}) with the derived
effective force (see Eq.~(\ref{B15})) for a study of the dynamics of
perturbations of the mean velocity. We show that under certain
conditions the large-scale instability can be excited which causes
formation of large-scale vortical structures.

\section{Linear velocity shear (Couette flow) in
homogeneous turbulence}

We consider a homogeneous turbulence with a mean linear velocity
shear, $ {\bf U}^{(s)} = (0, Sx, 0)$. This velocity field is a
steady state solution of the Navier-Stokes equation. Let us first
study the case when the velocity perturbations $\tilde{\bf U}(t, x,
z)$ are independent of $y$. The equations for the components
$\tilde{U}_x$ and $\tilde{U}_y$  of the velocity perturbations read
\begin{eqnarray}
\Big[{\partial \over \partial t} - \nu_{_{T}} \Delta \Big] \Delta \,
\tilde{U}_x &=& l_0^2 \, S \, \beta_0 \, \Delta \, \nabla_z^2 \,
\tilde{U}_y \;,
\label{TAA1}\\
\Delta \Big[{\partial \over \partial t} - \nu_{_{T}}  \, \Delta
\Big] \, \tilde{U}_y &=& - S  \, \Delta  \, \tilde{U}_x \;,
\label{TAA2}
\end{eqnarray}
and the component $\tilde{U}_z$ is determined by the continuity
equation $\bec{\nabla} {\bf \cdot} \tilde{\bf U} = 0$, where
$\beta_0=C_1 + C_2 - C_3 = C^2 \, (2 \mu^2 - 43 \, \mu + 63) / 315$.
In order to derive Eqs. (\ref{TAA1}) and (\ref{TAA2}) we calculate
$\bec{\nabla} {\bf \times} (\bec{\nabla} {\bf \times} \tilde{\bf
U})$ using Eq.~(\ref{B4}), that allows us to exclude the pressure
term from this equation. We also use Eq.~(\ref{B15}) for the
Reynolds stresses in the sheared turbulence. For simplicity, in
Eq.~(\ref{TAA2}) we neglect the small terms $\sim O[(l_0/L_S)^2]$,
where $L_S$ is the characteristic scale of the velocity shear.

We seek for a solution of Eqs. (\ref{TAA1}) and (\ref{TAA2}) in the
form
\begin{eqnarray}
\tilde{U}_{x,y} &=&  \exp(\gamma t) \, [A_{x,y} \, \cos(K_x \, x) +
B_{x,y} \, \cosh(K_z \, x)]
\nonumber\\
 &&\times \cos(K_z \, z + \phi) \;, \label{S1}
\end{eqnarray}
where the coefficients $A_{x,y}$, $B_{x,y}$, the angle $\phi$ and
the growth rate $\gamma$ of the instability are determined by the
boundary conditions. We choose the symmetric solution (relative the
point $x=0$), because the maximum growth rate of the symmetric mode
is higher than that of antisymmetric mode (see below). Perturbations
of the mean velocity grow in time due to the large-scale instability
with the growth rate
\begin{eqnarray}
\gamma = \sqrt{\beta_0} \, S \, l_0 \, K_z - \nu_{_{T}} (K_x^2 +
K_z^2) \; . \label{N20}
\end{eqnarray}
The maximum growth rate of perturbations of the mean velocity,
\begin{eqnarray}
\gamma_{\rm max} = {\beta_0 \, (S \, l_0)^2 \over 4 \nu_{_{T}}} -
\nu_{_{T}} \, K_x^2 \;, \label{S3}
\end{eqnarray}
is attained at $ K_z = K_m = \sqrt{\beta_0} \, S \, l_0 /2
\nu_{_{T}}$.

In order to determine the threshold required for the excitation of
the large-scale instability, we consider the solution of
Eqs.~(\ref{TAA1}) and~(\ref{TAA2}) with the following boundary
conditions for a layer of the thickness $L_S$ in the $x$ direction:
at $x=\pm \, L_S / 2$ the functions $\tilde{\bf U} = 0$ and
$\nabla_x \, (\tilde{U}_{x,y}) = 0$.  This yields the threshold
value of the wave number $K_x^{\rm cr}$, determined by the equation
\begin{eqnarray}
\tan(K_x^{\rm cr} L_S/ 2) = - \tanh(K_x^{\rm cr} L_S/2) \; .
\label{S2}
\end{eqnarray}
The condition $\gamma_{\rm max} > 0$ implies that $K_m \geq K_x^{\rm
cr}$. Therefore, the large-scale instability is excited when the
value of the shear $S$ exceeds the critical value $S_{\rm cr}$ that
is given by
\begin{eqnarray}
S_{\rm cr} \, \tau_0 = {2 \, K_x^{\rm cr} \, l_0 \over 3 \,
\sqrt{\beta_0}} \approx 4.7 \, {l_0 \over L_S} \;, \label{N21}
\end{eqnarray}
where $K_x^{\rm cr} = 2 \pi /L_S$. Note that the value of $K_x^{\rm
cr}$ for the the symmetric mode is smaller than that for
antisymmetric mode. This is the reason why the maximum growth rate
of the symmetric mode is larger than that of antisymmetric mode.

Note that the parameter $\beta_0$ depends on the scaling exponent
$\mu$ of the correlation time of the turbulent velocity field,
$\tau(k) \propto k^{-\mu}$. In particular, for the Kolmogorov
scaling, $\tau(k) \propto k^{-2/3}$, we arrive at $\beta_0 = 0.45$.
This case has been considered in \cite{EKR03}. The necessary
condition for the large-scale instability $(\beta_0 > 0)$ reads $2
\, \mu^2 - 43 \, \mu + 63 > 0 $, i.e., the instability is excited
when $0 \leq \mu < 1.58$ and $\mu > 19.9$. Note that the condition
$\mu > 19.9$ is not realistic. In the case of a turbulence with a
scale-independent correlation time, the exponent $\mu=0$ and the
parameter $\beta_0=0.2$.

For small hydrodynamic Reynolds numbers, the scaling of the
correlation time $\tau(k) \sim 1/(\nu k^2)$, i.e., $\mu = 2$, and
the parameter $\beta_0 < 0$. This implies that the instability of
the perturbations of the mean vorticity does not occur for small
Reynolds numbers in agreement with the recent results obtained in
\cite{RUK06} whereby an instability of the perturbations of the mean
vorticity in a random flow with large-scale velocity shear has not
been found using the second order correlation approximation and
assumption that the correlation time $\tau(k) \sim 1/(\nu k^2)$.
This approximation is valid only for small Reynolds numbers (see
discussion in \cite{RKL06}).

Let us consider now a more general case when the velocity
$\tilde{\bf U}$ depends on three spatial coordinates, i.e.,
$\tilde{\bf U}=\tilde{\bf U}(t, x, y, z)$. The equations for the
components $\tilde{U}_x$ and $\tilde{U}_y$  of the velocity
perturbations read
\begin{eqnarray}
\bigg({\partial \over \partial \,t} &+& U^{(s)} \, \nabla_y -
\nu_{_T} \Delta \bigg) \, \Delta \tilde{U}_x = l_0^2\,S\, \,
\Delta\, \big[\beta_0 \, \Delta_{H} \tilde{U}_y
\nonumber \\
&+& (\beta_1 - \beta_2) \, \nabla_x\,\nabla_y \tilde{U}_x\big]\;,
\label{AA1}\\
\Delta\bigg({\partial \over \partial \,t} &+& U^{(s)} \, \nabla_y -
\nu_{_T} \Delta\bigg)\tilde{U}_y = l_0^2\,S \, \Delta \,
\big[\beta_2 \, (\Delta - \nabla_y^2) \,\tilde{U}_x
\nonumber \\
&+& (\beta_1 - \beta_0) \, \nabla_x \nabla_y \, \tilde{U}_y \big] +
S \big(2 \nabla_y^2 - \Delta \big) \, \tilde{U}_x \;, \label{AA2}
\end{eqnarray}
and the component $\tilde{U}_z$ is determined by the continuity
equation $\bec{\nabla} {\bf \cdot} \tilde{\bf U} = 0$. Here
$\Delta_{H} = \Delta - \nabla_x^2$, $\, \beta_{1} = 2 C_1 - C_2 =
C^2 \, (8 \, \mu^2 - 39 \, \mu + 231) / 630$ and $\beta_{2} = C_1 +
C_3 = C^2 \, (4 \, \mu^2 - 51 \, \mu + 91) / 630$. In order to
derive Eqs.~(\ref{AA1}) and (\ref{AA2}) we calculate $\bec{\nabla}
{\bf \times} (\bec{\nabla} {\bf \times} \tilde{\bf U})$ using
Eq.~(\ref{B4}), that allows us to exclude the pressure term from
this equation. For the derivation of Eqs.~(\ref{AA1}) and
(\ref{AA2}) we also use Eq.~(\ref{B15}) for the Reynolds stresses in
the sheared turbulence. Equations~(\ref{AA1}) and (\ref{AA2}) can be
reduced to the Orr-Sommerfeld equation if we replace $\nu_{T}$ by
$\nu$ and set $\beta_n=0$ (see, e.g., \cite{DR81,SH01,CJJ03}, and
references therein).

\begin{figure}
\vspace*{2mm} \centering
\includegraphics[width=7cm]{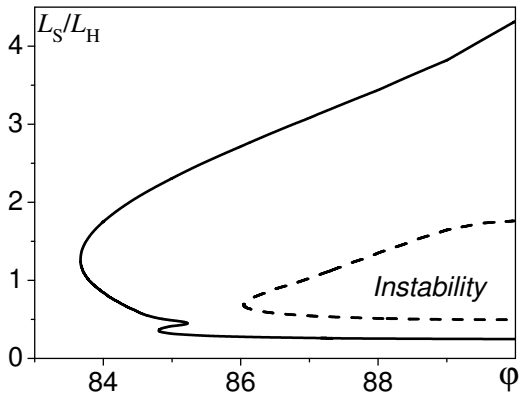}
\caption{\label{FIG-1} Range of parameters ($L_S / L_{H}$;
$\varphi$) for which the large-scale instability occurs  for Couette
background flow and for different values of the large-scale shear:
$S \, \tau_0 = 0.2$ (dashed line), $S \, \tau_0 = 0.4$ (solid line).
Here $L_S / l_0 = 30$.}
\end{figure}

\begin{figure}
\vspace*{2mm} \centering
\includegraphics[width=7cm]{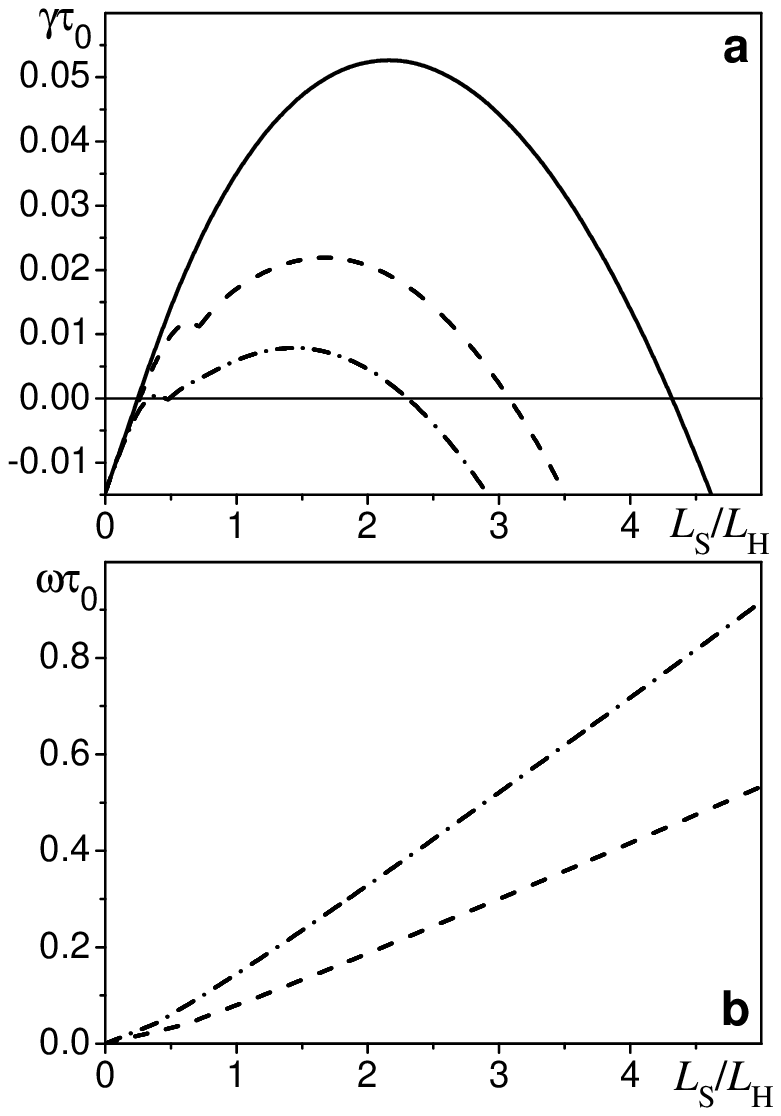}
\caption{\label{FIG-2} The growth rate (a) of the large-scale
instability $\gamma \, \tau_0\, $ and frequencies $\omega \,
\tau_0\, $ of the generated modes (b) versus $L_S / L_{H}$ for
Couette background flow and for different angles $\varphi$: $\,
\varphi = 85^\circ$ (dashed-dotted line), $\varphi = 87^\circ$
(dashed line), $\varphi = 90^\circ$ (solid line). Here $S \, \tau_0
= 0.4$, $\, L_S / l_0 = 30$ and $\omega(\varphi = 90^\circ)=0 $.}
\end{figure}

\begin{figure}
\vspace*{2mm} \centering
\includegraphics[width=7cm]{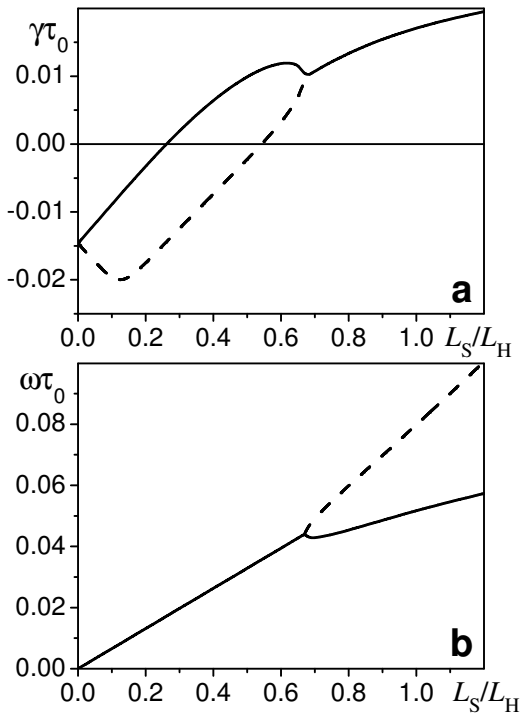}
\caption{\label{FIG-3} The growth rate $\gamma \, \tau_0\,$ (a) and
the frequency $\omega \, \tau_0\,$ (b) versus $L_S / L_{H}$ of the
first (solid line) and the second (dashed line) modes which have the
highest growth rates for Couette background flow. Here the angle
$\varphi = 87^{\circ}$, $\, S \, \tau_0 = 0.4$ and $L_S / l_0 =
30$.}
\end{figure}

\begin{figure}
\vspace*{2mm} \centering
\includegraphics[width=7cm]{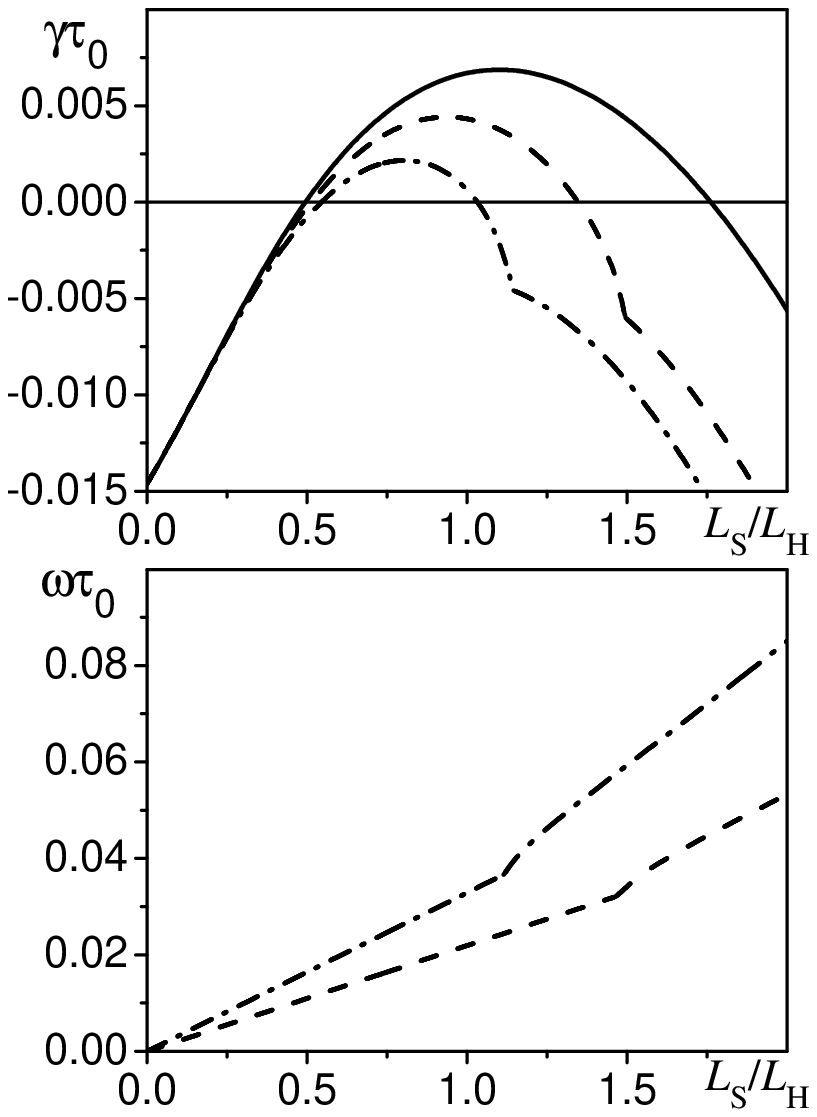}
\caption{\label{FIG-4} The growth rate (a) of the large-scale
instability $\gamma \, \tau_0\, $ and frequencies $\omega \,
\tau_0\, $ of the generated modes (b) versus $L_S / L_{H}$ for
Couette background flow and for different angles $\varphi$:  $\,
\varphi = 87^\circ$ (dashed-dotted line), $\varphi = 88^\circ$
(dashed line), $\varphi = 90^\circ$ (solid line). Here $S \, \tau_0
= 0.2$, $\, L_S / l_0 = 30$ and $\omega(\varphi = 90^\circ)=0$.}
\end{figure}

\begin{figure}
\vspace*{2mm} \centering
\includegraphics[width=7cm]{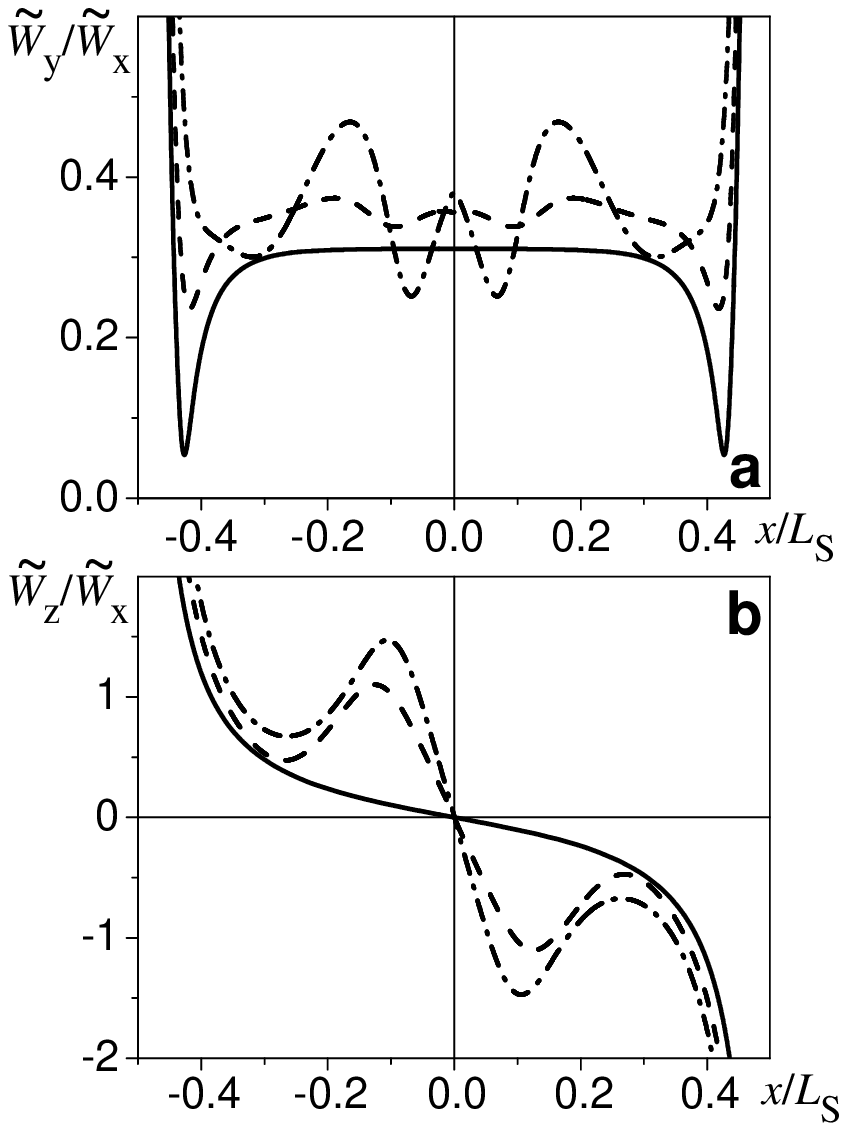}
\caption{\label{FIG-5} The spatial profiles of the ratios of
vorticity components $\tilde W_y/\tilde W_x$ (a) and $\tilde W_z/
\tilde W_x$ (b) for modes with the maximum growth rates of the
large-scale instability in Couette background flow and for different
angles $\varphi$: $\, \varphi = 85^\circ$ (dashed-dotted line),
$\varphi = 87^\circ$ (dashed line), $\varphi = 90^\circ$ (solid
line). Here $S \, \tau_0 = 0.4$ and $L_S / l_0 = 30$.}
\end{figure}

We seek for a solution of Eqs.~(\ref{AA1})-(\ref{AA2}) in the form
$\propto \Psi(x) \, \exp(\gamma t + i \, \omega t + i \, {\bf K}_H
\, {\bf \cdot \, r})$, where ${\bf K}_H$ is the wave number that is
perpendicular to the $x$-axis. After the substitution of this
solution into Eqs.~(\ref{AA1})-(\ref{AA2}) we obtain the system of
the ordinary differential equations which is solved numerically. We
consider the solution of Eqs.~(\ref{AA1})-(\ref{AA2}) with the
following boundary conditions for a layer of the thickness $L_S$ in
the $x$ direction: at $x=\pm \, L_S / 2$ the functions $\tilde{\bf
U} = 0$ and $\nabla_x \, (\tilde{U}_{x,y}) = 0$. These boundary
conditions with a linear velocity shear corresponds to the Couette
flow.

In this Section we show that in a small-scale turbulence the
large-scale Couette flow can be unstable under certain conditions.
The range of parameters ($L_S / L_{H}$; $\varphi$) for which the
large-scale instability occurs is shown in Fig.~\ref{FIG-1}, where
$L_H= 2 \pi /K_H$, $\, K_H = (K_y^2 + K_z^2)^{1/2}$ and $\varphi$ is
the angle between the wave vector ${\bf K}_H$ and the direction of
the mean sheared velocity ${\bf U}^{(s)}$. In
Figs.~\ref{FIG-2}-\ref{FIG-4} we show the growth rate of the
large-scale instability $\gamma \, \tau_0$ and the frequencies of
the generated modes $\omega \, \tau_0$ versus $L_S / L_{H}$. The
growth rates of the large-scale instability increase with the
increase of the angle $\varphi$, while the frequencies of the
generated modes decrease with the angle $\varphi$ so that
$\omega(\varphi \to 90^\circ) \to 0$. The growth rate of the
large-scale instability reaches the maximum value at $\varphi =
90^\circ$. In addition, the range of angles $\varphi$ for which the
large-scale instability occurs, is small and located in the vicinity
of $\varphi = 90^\circ$ (see Fig.~\ref{FIG-1}). Therefore, $K_y \ll
K_z$ and since $L_z \sim L_S$, the size of the structures in the
direction of ${\bf U}^{(s)}$ is much larger than the sizes of the
structures along $x$ and $z$ directions. This implies that the
large-scale structures formed due to this instability are stretched
along the mean sheared velocity ${\bf U}^{(s)}$.

The curves in Figs.~\ref{FIG-2}-\ref{FIG-4} have a point $L_{\ast}$
whereby the first derivative of the growth rate of the large-scale
instability with respect to the wave number $K_H$ has a singularity.
At this point there is a bifurcation which is illustrated in
Fig.~\ref{FIG-3}. In particular, the growth rates and the
frequencies for the first and the second modes which have the
highest growth rates are shown in Fig.~\ref{FIG-3}a
and~\ref{FIG-3}b. When the size of perturbations $L_{H} < L_{\ast}$,
the frequencies of the first and the second modes are different, but
the growth rates are the same. Therefore, at the point $L_{H} =
L_{\ast}$, there is a generation of two different modes with the
same growth rate.  On the other hand, when the size of perturbations
$L_{H} > L_{\ast}$, the growth rates of the first and the second
modes are different, but the frequencies are the same.

The maximum growth rate of perturbations of the mean velocity, $
\gamma_{\rm max}$, is attained at $ K_H = K_m$, and the value $K_m$
increases with the increase of the angle $\varphi$ between the wave
vector ${\bf K}_H$ and the direction of the mean sheared velocity
${\bf U}^{(s)}$. The increase of shear $S$ promotes the large-scale
instability, i.e., it cause the increase of the range for the
instability (see Fig.~\ref{FIG-1}) and the maximum growth rate (see
Figs.~\ref{FIG-2} and \ref{FIG-4}). The characteristic spatial scale
$L_m = 2 \pi / K_m$ and the time scale $t_{\rm inst} \sim
\gamma_{\rm max}^{-1}$ for the instability are much larger than the
characteristic turbulent scales. This justifies separation of scales
which is required for the validity of the mean-field theory applied
in the present study. The spatial profiles of the ratios of
vorticity components $\tilde W_y/\tilde W_x$ and $\tilde W_z/\tilde
W_x$ for perturbations in Couette background flow are shown in
Fig.~\ref{FIG-5}. The function $\tilde W_y/\tilde W_x$ is symmetric
relative to the center of the flow at $x=0$, while the function
$\tilde W_z/\tilde W_x$ is antisymmetric. Since the function $\tilde
W_x \to 0$ at the boundaries of the flow, the ratios of vorticity
components $\tilde W_y/\tilde W_x$ and $\tilde W_z/\tilde W_x$ tend
to $\to \pm \, \infty$ at the boundaries.

The numerical results for the case $\varphi = 90^\circ$ shown in
Figs.~\ref{FIG-2}, \ref{FIG-4} and~\ref{FIG-5} coincide with the
analytical predictions based on Eqs.~(\ref{S1})-(\ref{N21}). For
instance, the threshold value of the shear at $L_S / l_0 = 30$ is
$S_{\rm cr} \, \tau_0 \approx 0.157$ in agreement with
Eq.~(\ref{N21}). The ratio of vorticity components $\tilde
W_y/\tilde W_x \approx 0.3$ at $x=0$ for modes with the maximum
growth rate of the large-scale instability. This is in agreement
with this ratio of $\tilde W_y/\tilde W_x$ obtained using
Eq.~(\ref{S1}). The maximum growth rates of perturbations of the
mean velocity are in agreement with Eqs.~(\ref{S3}) and (\ref{S2}).
When we switch off the turbulence, the large-scale instability does
not excited, etc.

The growing modes with a nonzero frequency discussed in this Section
can be regarded as the turbulent analogue of the
Tollmien-Schlichting waves. In laminar flows the
Tollmien-Schlichting waves are growing solutions of the
Orr-Sommerfeld equation and the molecular viscosity promotes the
excitation of the Tollmien-Schlichting waves (see, e.g.,
\cite{SH01}). On the other hand, the turbulent Tollmien-Schlichting
waves are excited by a small-scale sheared turbulence, i.e., by a
combined effect of the turbulent Reynolds stress-induced generation
of perturbations of the mean vorticity and the background sheared
motions.

\section{Quadratic velocity shear (Poiseuille flow) in homogeneous turbulence}

Now we consider a homogeneous turbulence with an imposed large-scale
quadratic  velocity shear, ${\bf U}^{(s)} = S_\ast \, x \, (1-x/L_S)
\, {\bf e}_y$. The equations for the components $\tilde{U}_x$ and
$\tilde{U}_y$  of the velocity perturbations read
\begin{eqnarray}
&&\bigg({\partial \over \partial \,t} + U^{(s)} \, \nabla_y -
\nu_{_T} \Delta \bigg)\Delta \tilde{U}_x = l_0^2\,S\, \Delta\,
\Big(\beta_0\, \Delta_{H} \tilde{U}_y
\nonumber \\
&&\qquad + (\beta_1 - \beta_2) \, \nabla_x\,\nabla_y \, \tilde{U}_x
\Big) + S' \,  \nabla_y \, \tilde{U}_x\, ,
\label{AB1}\\
&&\Delta\bigg({\partial \over \partial \,t} + U^{(s)} \, \nabla_y -
\nu_{_T} \Delta\bigg) \, \tilde{U}_y = S \big(2 \nabla_y^2 - \Delta
\big) \, \tilde{U}_x
\nonumber \\
&&\qquad - 2 \, S' \, \nabla_x\, \tilde{U}_x + l_0^2\,S \, \Delta \,
\Big[(\beta_1 - \beta_0) \, \nabla_x \nabla_y \, \tilde{U}_y
\nonumber \\
&&\qquad + \beta_2 \, (\Delta - \nabla_y^2) \,\tilde{U}_x\Big] +
l_0^2 \, S' \, \Big[2 \beta_1 \nabla_x \nabla_y (\nabla_x
\tilde{U}_y
\nonumber \\
&&\qquad - \nabla_y \tilde{U}_x) + \Delta \, \big[( 2 \beta_2 +
\beta_1) \, \nabla_x \tilde{U}_x
\nonumber \\
&&\qquad + (\beta_2 - \beta_0) \, \nabla_y \tilde{U}_y \big] \Big]
\;, \label{AB2}
\end{eqnarray}
and the component $\tilde{U}_z$ is determined by the continuity
equation $\bec{\nabla} {\bf \cdot} \tilde{\bf U} = 0$, where $S(x) =
\nabla_x\, U^{(s)}$ and $S'= \nabla_x\, S$. In order to derive
Eqs.~(\ref{AB1}) and (\ref{AB2}) we calculate $\bec{\nabla} {\bf
\times} (\bec{\nabla} {\bf \times} \tilde{\bf U})$ using
Eq.~(\ref{B4}). We seek for a solution of Eqs.~(\ref{AB1}) and
(\ref{AB2}) in the form $\propto \Psi(x) \, \exp(\gamma t + i \,
\omega t + i \, {\bf K}_H \, {\bf \cdot \, r})$, where ${\bf K}_H$
is the wave number that is perpendicular to the $x$-axis. After the
substitution of this solution into Eqs.~(\ref{AB1}) and (\ref{AB2})
we obtain the system of the ordinary differential equations which is
solved numerically. We consider the solution of
Eqs.~(\ref{AB1})-(\ref{AB2}) with the following boundary conditions
for a layer of the thickness $L_S$ in the $x$ direction: at $x=\pm
\, L_S / 2$ the functions $\tilde{\bf U} = 0$ and $\nabla_x \,
(\tilde{U}_{x,y}) = 0$. These boundary conditions with a quadratic
large-scale velocity shear corresponds to the Poiseuille flow. We
show below that in a small-scale turbulence the large-scale
Poiseuille flow can be unstable with respect to small perturbations.

\begin{figure}
\vspace*{2mm} \centering
\includegraphics[width=7cm]{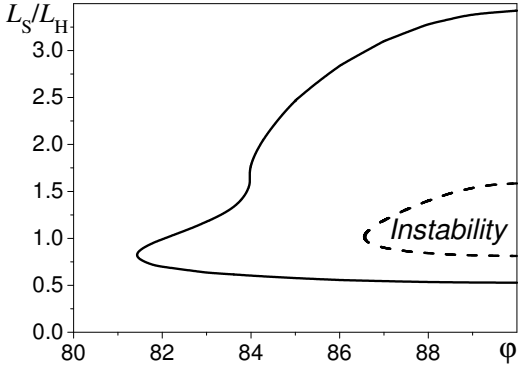}
\caption{\label{FIG-6} Range of parameters ($L_S / L_{H}$;
$\varphi$) for which the large-scale instability for Poiseuille
background flow occurs, and for different values of the large-scale
shear: $S_\ast \, \tau_0 = 0.5$ (dashed line) and $S_\ast \, \tau_0
= 0.6$ (solid line). Here $L_S / l_0 = 30$ and $S_\ast=S(x=0)$.}
\end{figure}

\begin{figure}
\vspace*{2mm} \centering
\includegraphics[width=7cm]{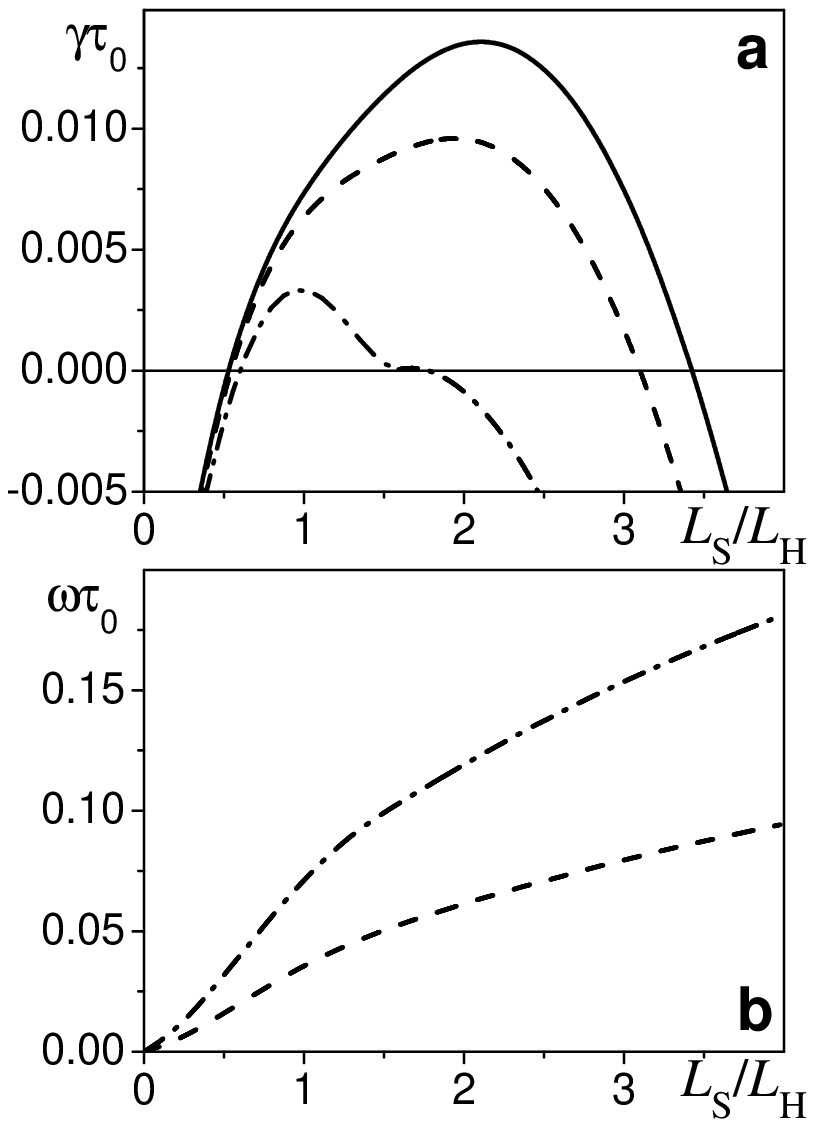}
\caption{\label{FIG-7} The growth rate (a) of the large-scale
instability $\gamma \, \tau_0\, $ and frequencies $\omega \,
\tau_0\, $ of the generated modes (b) versus $L_S / L_{H}$ for
Poiseuille background flow  and  for different angles $\varphi$: $\,
\varphi = 84^\circ$ (dashed-dotted line), $\varphi = 87^\circ$
(dashed line), $\varphi = 90^\circ$ (solid line). Here $S_\ast \,
\tau_0 = 0.6$, $\, L_S / l_0 = 30$ and $\omega(\varphi = 90^\circ)=0
$.}
\end{figure}

\begin{figure}
\vspace*{2mm} \centering
\includegraphics[width=7cm]{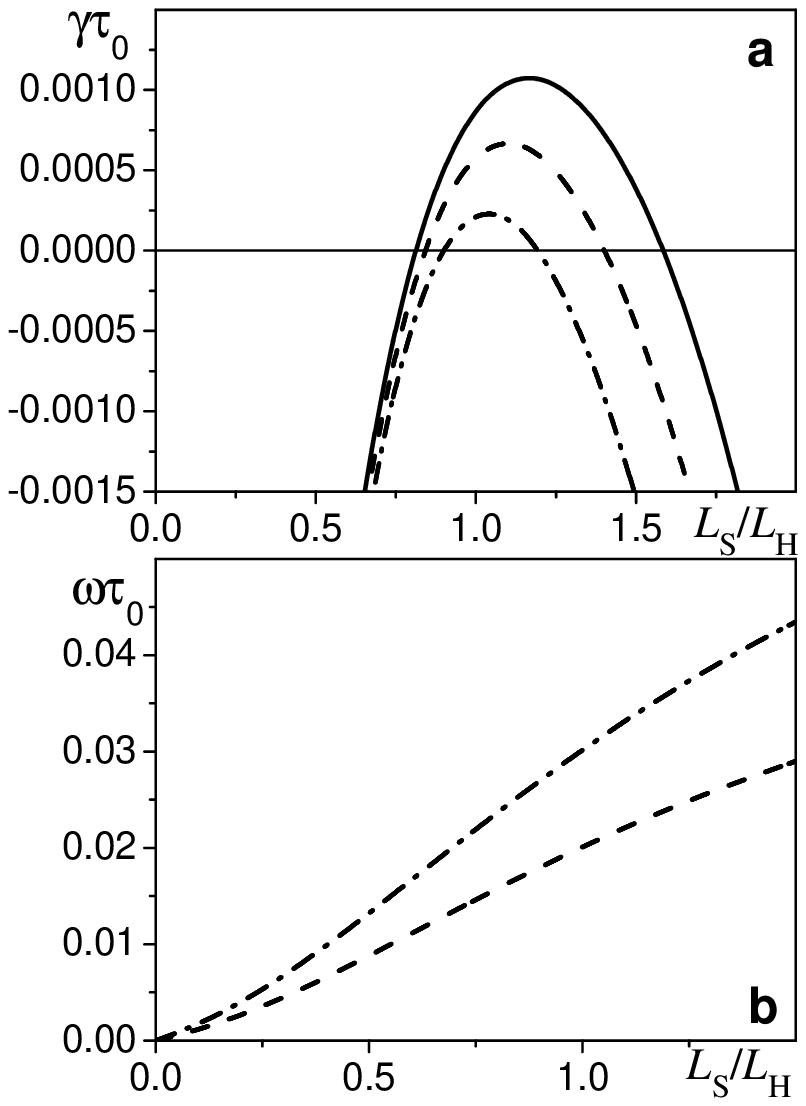}
\caption{\label{FIG-8} The growth rate (a) of the large-scale
instability $\gamma \, \tau_0\, $ and frequencies $\omega \,
\tau_0\, $ of the generated modes (b) versus $L_S / L_{H}$ for
Poiseuille background flow and for different angles $\varphi$: $\,
\varphi = 87^\circ$ (dashed-dotted line), $\varphi = 88^\circ$
(dashed line), $\varphi = 90^\circ$ (solid line). Here $S_\ast \,
\tau_0 = 0.5$, $\, L_S / l_0 = 30$ and $\omega(\varphi = 90^\circ)=0
$.}
\end{figure}

\begin{figure}
\vspace*{2mm} \centering
\includegraphics[width=7cm]{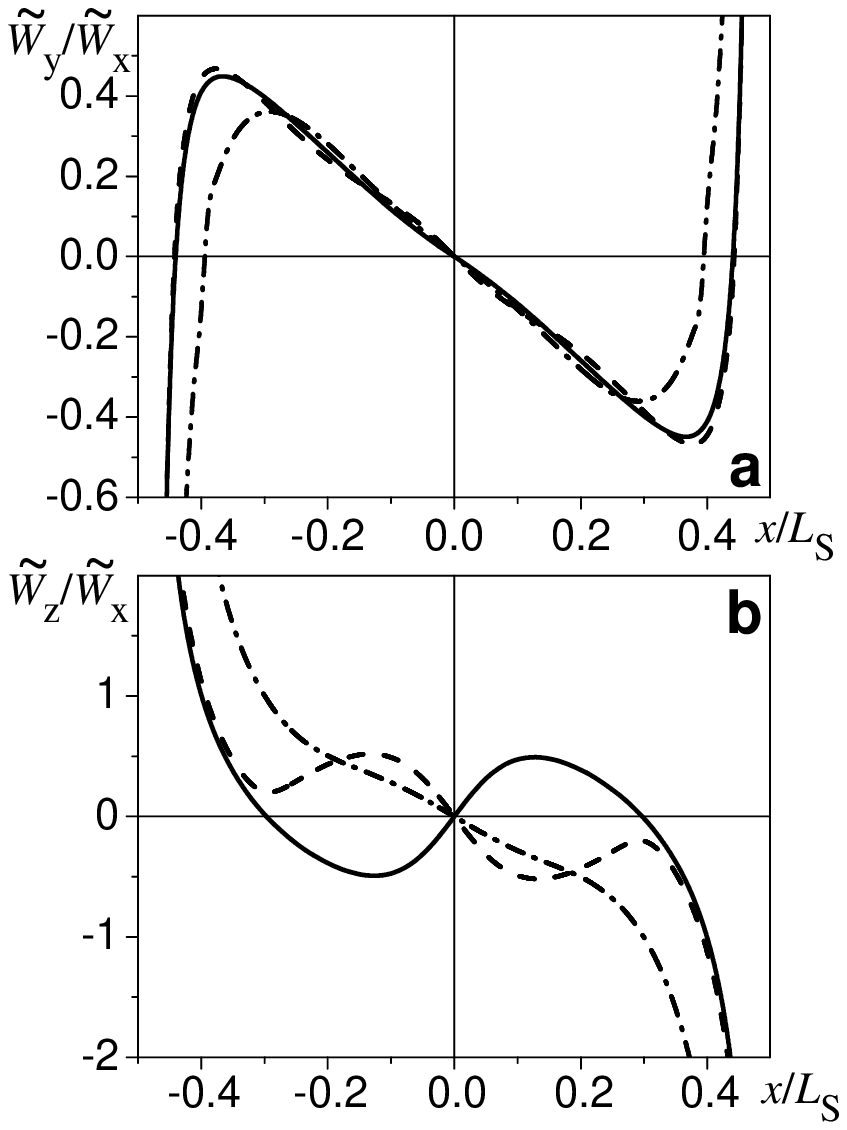}
\caption{\label{FIG-9} The spatial profiles of the ratios of
vorticity components $\tilde W_y/\tilde W_x$ (a) and $\tilde
W_z/\tilde W_x$ (b) for modes with the maximum growth rates of the
large-scale instability in Poiseuille background flow and for
different angles $\varphi$: $\, \varphi = 84^\circ$ (dashed-dotted
line), $\varphi = 87^\circ$ (dashed line), $\varphi = 90^\circ$
(solid line). Here $S_\ast \, \tau_0 = 0.6$ and $L_S / l_0 = 30$.}
\end{figure}

The range of parameters ($L_S / L_{H}$; $\varphi$) for which the
large-scale instability in the Poiseuille background flow occurs is
shown in Fig.~\ref{FIG-6} for different values of the large-scale
shear, where $S_\ast=S(x=0)$. The growth rates of this instability
and the frequencies of the generated turbulent Tollmien-Schlichting
waves are shown in Figs.~\ref{FIG-7} and \ref{FIG-8}. The spatial
profiles of the ratios of vorticity components $\tilde W_y/\tilde
W_x$ and $\tilde W_z/\tilde W_x$ in Poiseuille background flow for
modes with the maximum growth rates of the large-scale instability
are shown in Fig.~\ref{FIG-9}. The general behaviour of the
large-scale instability in the Poiseuille background flow is similar
to that for the Couette background flow. In particular, the growth
rates of the large-scale instability increase with the increase of
the angle $\varphi$ between the wave vector ${\bf K}_H$ and the
direction of the mean sheared velocity ${\bf U}^{(s)}$, reaching the
maximum value at $\varphi = 90^\circ$. The frequencies $\omega \,
\tau_0\, $ of the generated turbulent Tollmien-Schlichting waves by
the large-scale instability decrease with the increase of the angle
$\varphi$ and $\omega \to 0$ at $\varphi \to 90^\circ$. The values
$K_m$ at which the growth rates of the large-scale instability reach
the maximum values increase with the increase of the angle
$\varphi$. The range for the large-scale instability and the growth
rates of perturbations in the Poiseuille background flow increases
with the increase of shear. This implies that increase of shear
promotes the large-scale instability.

For the Poiseuille flow the large-scale instability can be excited
for smaller angles $\varphi$ than that for the Couette background
flow. On the other hand, the thresholds for the instability in the
value of shear and in the value of $L_S / L_{H}$ for Poiseuille
background flow are larger than that for the Couette background
flow. A difference between the Couette and Poiseuille background
flows can be also seen in Figs.~\ref{FIG-5} and~\ref{FIG-9} for the
spatial profiles of the ratios of vorticity components $\tilde
W_y/\tilde W_x$ and $\tilde W_z/\tilde W_x$. This difference is
caused by the different geometries in these flows. In particular,
the first spatial derivatives of the flow velocity in the Poiseuille
background flow are antisymmetric relative to the center of the flow
at $x=0$, while they are symmetric (constant) in the Couette
background flow. This is the reason of that the spatial profile of
$\tilde W_y/\tilde W_x$ is symmetric relative to $x=0$ in the
Couette background flow, and it is antisymmetric in the Poiseuille
flow.

\section{Nonuniform velocity shear in inhomogeneous turbulence}

In this Section we consider a more complicated form of nonuniform
velocity shear in an inhomogeneous turbulence. For simplicity we
consider the case when the small perturbations of the mean velocity
$\tilde{\bf U}$ are independent of $y$. The equations for the
components $\tilde{U}_x$ and $\tilde{U}_y$  of the velocity
perturbations in an inhomogeneous turbulence with a nonuniform shear
read
\begin{eqnarray}
\Delta \Big[{\partial \over \partial t} &-& \nu_{_{T}}  \, \Delta
\Big] \, \tilde{U}_x  = \beta_0 \, \Big[l_0^2  \, S \, \Delta -
\nabla_x^2 \, \big(l_0^2 \, S\big) \Big] \, \nabla_z^2 \, \tilde
U_y
\nonumber\\
&-& 2  \, \Big(\nabla_x^2 \, \nu_{_{T}}\Big) \, \nabla_z^2 \,
\tilde{U}_x  \;,
\label{A1}\\
\Big[{\partial \over \partial t} &-& \nu_{_{T}} \Delta \Big] \,
\tilde{U}_y = \Big[-S + \beta_{1} \, \nabla_x \big(l_0^2 \, S \,
\big) \, \nabla_x
\nonumber\\
&+&  \beta_{2} \, l_0^2 \, S \, \Delta \Big]  \, \tilde{U}_x +
\Big(\nabla_x \, \nu_{_{T}} \Big) \, \nabla_x  \, \tilde{U}_y \;,
\label{A2}
\end{eqnarray}
and the component $\tilde{U}_z$ is determined by the continuity
equation $\bec{\nabla} {\bf \cdot} \tilde{\bf U} = 0$, where $S(x) =
\nabla_x\, U^{(s)}$. Equation~(\ref{A2}) is the $y$ component of
Eq.~(\ref{B4}) with $\nabla_y \tilde P =0$, while Eq.~(\ref{A1}) is
the $x$ component of $\bec{\nabla} {\bf \times} (\bec{\nabla} {\bf
\times} \tilde{\bf U})$ determined from Eq.~(\ref{B4}). We consider
the solution of Eqs.~(\ref{A1}) and (\ref{A2}) with the following
boundary conditions  for a layer of the thickness $L_S$ in the $x$
direction: at $x=\pm \, L_S/2$ the functions $\tilde{\bf U} = 0$ and
$\nabla_x \, (\tilde{U}_{x,y}) = 0$.

\begin{figure}
\vspace*{2mm} \centering
\includegraphics[width=7cm]{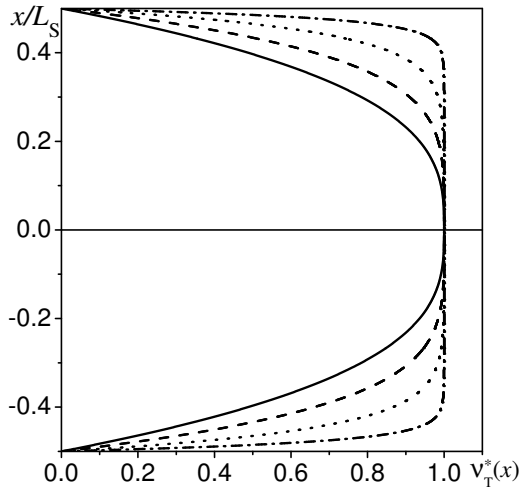}
\caption{\label{FIG-10} The spatial profile of the normalized
turbulent viscosity $\nu_{_{T}}^\ast(x)$ for different values of the
parameter $\alpha$: $\; \alpha=6$ (solid line), $\alpha=10$ (dashed
line), $\alpha=20$ (dotted line), $\alpha=50$ (dashed-dotted line).}
\end{figure}

\begin{figure}
\vspace*{2mm} \centering
\includegraphics[width=7cm]{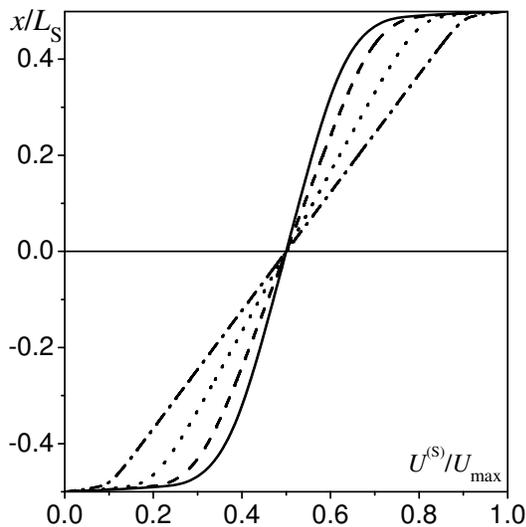}
\caption{\label{FIG-11} The mean velocity profile $U^{(s)}(x)/U_{\rm
max}$ for different values of the parameter $\alpha$: $\; \alpha=6$
(solid line), $\alpha=10$ (dashed line), $\alpha=20$ (dotted line),
$\alpha=50$ (dashed-dotted line), where $U_{\rm max} = u_{\star} /
\kappa$.}
\end{figure}

\begin{figure}
\vspace*{2mm} \centering
\includegraphics[width=7cm]{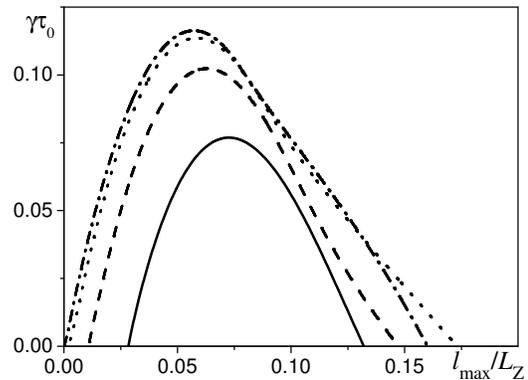}
\caption{\label{FIG-12} The growth rate of the large-scale
instability versus $l_{\rm max} / L_z$ in inhomogeneous turbulence
with nonuniform velocity shear for different values of the parameter
$\alpha$: $\; \alpha = 6$ (solid line), $\alpha = 10$ (dashed line),
$\alpha = 20$ (dotted line) and $\alpha = 50$ (dashed-dotted line).
Here $l_{\rm max} = l_0(x \to 0.5 \,L_z)$.}
\end{figure}

We consider a "log-linear" velocity profile for the background
large-scale flow in an inhomogeneous turbulence. In particular, we
use the following relationship for the velocity shear $S(x) =
u_{\star}^2 / \nu_{_{T}}(x)$ and the eddy viscosity $\nu_{_{T}}(x) =
u_{\star} \, l_0(x)$, where $l_0(x) = \kappa \, \eta(x) \, L_S $ is
the turbulence length scale, $\kappa$ is the von K\'arm\'an
constant, $u_{\star}$ is the friction velocity, $\eta(x)$ is the
dimensionless function that characterizes the spatial profile of the
background velocity shear and inhomogeneity of small-scale
turbulence (see below). These relationships are usually used for the
logarithmic boundary layer profiles (see, e.g., \cite{MY75}). The
spatial profile $\eta(x)$ for $0 \leq x \leq L_S/2$ is chosen in the
form
\begin{eqnarray}
\eta(x) = a_1 \,\big[1-\exp(-a_0\, \tilde x)\big] + a_2\, \tilde x +
a_3\, \tilde x^2 + a_4\, \tilde x^3 \;,
\nonumber\\
\label{B30}
\end{eqnarray}
where $\tilde x= x/L_S-1/2$, the coefficients $a_k$ are determined
by the following conditions: at $x=0$ the functions $\eta = 1$, $\,
\nabla_x \eta = 0$, $\, \nabla^2_x \eta = 0$, $\, \nabla^3_x \eta =
0$, and at $x=-L_S/2$ the derivative $\nabla_x \eta = \alpha / L_S$.
Here $\alpha$ is a free parameter that characterizes the
inhomogeneities of small-scale turbulence. The spatial profile of
the normalized turbulent viscosity $\nu_{_{T}}^\ast(x) =
\nu_{_{T}}(x) / (\kappa \, u_{\star} \, L_S) \equiv \eta(x)$ is
shown in Fig.~\ref{FIG-10} for different values of the parameter
$\alpha$. The function $\nu_{_{T}}^\ast(x)$ is chosen to be
symmetric relative the point $x=0$. The minimum possible value of
the parameter $\alpha$ is $\alpha = 6$. We have chosen the velocity
shear profile $U^{(s)}(x)$ so that the logarithmic velocity profile
near the boundaries can be matched with the linear shear velocity
for the central part of the background flow. Such kind of flow is
typical for the atmospheric boundary layer. Figure~\ref{FIG-11}
shows the mean velocity profile $U^{(s)}(x)/U_{\rm max}$  for
different values of the parameter $\alpha$, where $U_{\rm max} =
u_{\star} / \kappa$.

\begin{figure}
\vspace*{2mm} \centering
\includegraphics[width=7cm]{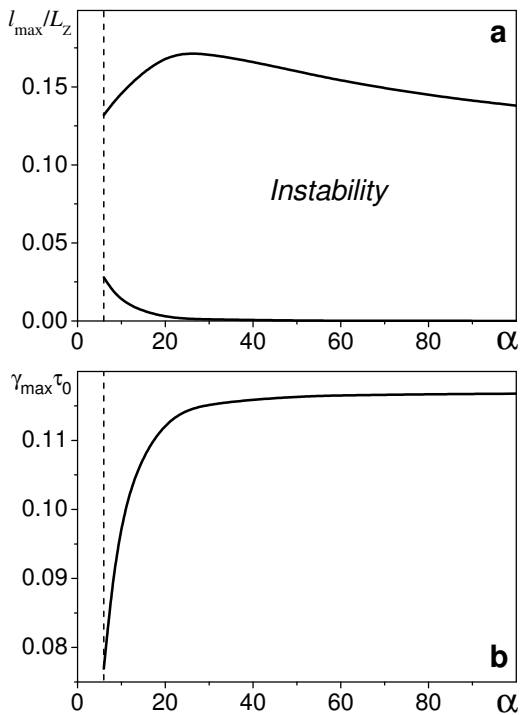}
\caption{\label{FIG-13}  (a). The range of parameters ($l_{\rm max}
/ L_z$; $\alpha$) for which the large-scale instability in
inhomogeneous turbulence with nonuniform velocity shear occurs. (b).
The maximum growth rate $\gamma_{\rm max} \, \tau_0$ of the
large-scale instability versus the parameter $\alpha$. Here $l_{\rm
max} = l_0(x \to 0.5 \,L_z)$.}
\end{figure}

\begin{figure}
\vspace*{2mm} \centering
\includegraphics[width=7cm]{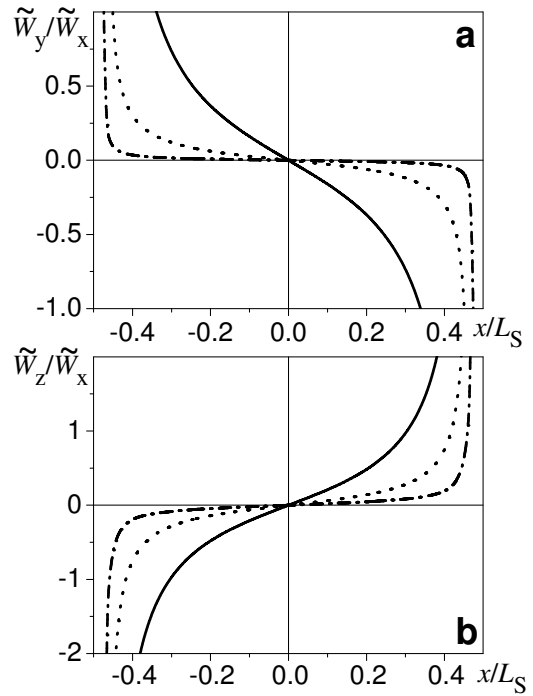}
\caption{\label{FIG-14} The spatial profiles of the ratios of
vorticity components $\tilde W_y/\tilde W_x$ (a) and $\tilde
W_z/\tilde W_x$ (b) for modes with the maximum growth rates of the
large-scale instability in inhomogeneous turbulence with nonuniform
velocity shear for different values of the parameter $\alpha$: $\;
\alpha = 6$ (solid line), $\alpha = 20$ (dashed line) and $\alpha =
50$ (dashed-dotted line). Here $l_{\rm max} = l_0(x \to 0.5
\,L_z)$.}
\end{figure}

We seek for a solution of Eqs.~(\ref{A1}) and (\ref{A2}) in the form
$\propto \Psi(x) \, \exp(\gamma t + i K_z \, z)$. After the
substitution of this solution into Eqs.~(\ref{A1}) and (\ref{A2}) we
obtain the system of the ordinary differential equations which is
solved numerically. The growth rate $\gamma \tau_0$ of the
large-scale instability versus $l_{\rm max} / L_z$ is shown in
Fig.~\ref{FIG-12}, where $L_z= 2 \pi /K_z$ is the size of
perturbations in $z$ direction and $l_{\rm max} = \kappa \, L_S$ is
the maximum value of the turbulent length scale $l_0$ when $\eta \to
1$ $\, (x \to 1)$. The range of parameters ($l_{\rm max} / L_z$;
$\alpha$) for which the large-scale instability occurs is shown in
Fig.~\ref{FIG-13}a. The vertical dashed line in Fig.~\ref{FIG-13}
indicates that the minimum possible value of the parameter $\alpha$
is $\alpha_{\rm min} = 6$. Figure~\ref{FIG-13}b demonstrates that
the increase of the parameter $\alpha$ causes the increase of the
maximum growth rate of the large-scale instability. The growth rate
of the large-scale instability for the inhomogeneous turbulence with
a large-scale nonuniform shear is much larger than that for the
Couette and Poiseuille background flows.

The spatial profiles of the ratios of vorticity components $\tilde
W_y/\tilde W_x$ and $\tilde W_z/\tilde W_x$ for modes with the
maximum growth rates of the large-scale instability are shown in
Fig.~\ref{FIG-14}. These profiles are different from that for the
Couette and Poiseuille background flows. The components $\tilde W_y$
and $\tilde W_z$ of perturbations of the mean vorticity in the
central part of the flow are usually much smaller than the component
$\tilde W_x$. Inspection of Figs.~\ref{FIG-12} and~\ref{FIG-13}a
shows that the parameter $l_{\rm max} / L_z < 0.17$. The
characteristic time scale for the instability is much larger than
the characteristic turbulent time. This justifies separation of
scales which is required for the validity of the mean-field theory
used here.

Note that in the interval $-L_S / 2 \leq x \leq 0$ the obtained
results discussed in this Section imply a stability theory for the
turbulent boundary layer. Our study shows that the turbulent
boundary layer can be unstable under certain conditions.

\section{Discussion}

In this study the theoretical approach proposed in \cite{EKR03} is
further developed and applied to investigate the large-scale
instability in a nonhelical turbulence with a nonuniform shear and a
more general form of the perturbations of the mean vorticity. In
particular, we consider three types of the background large-scale
sheared flows imposed on small-scale turbulence: Couette flow
(linear velocity shear) and Poiseuille flow (quadratic velocity
shear) in a small-scale homogeneous turbulence, and a more
complicated nonuniform velocity shear with the logarithmic velocity
profile near the boundaries matched with the linear shear velocity
for the central part of the background flow. This nonuniform
velocity shear is imposed on an inhomogeneous turbulence. The latter
flow is typical for the atmospheric boundary layer.

We show that the large-scale Couette and Poiseuille flows imposed on
a small-scale turbulence are unstable with respect to small
perturbations due to the excitation of the large-scale instability.
This instability causes generation of large-scale vorticity and
formation of large-scale vortical structures. The size of the formed
vortical structures in the direction of the background velocity
shear is much larger than the sizes of the structures in the
directions perpendicular to the velocity shear. Therefore, the
large-scale structures formed during this instability are stretched
along the mean sheared velocity. Increase of shear promotes the
large-scale instability. The thresholds for the excitation of the
large-scale instability in the value of shear and the aspect ratio
of structures for Poiseuille background flow are larger than that
for the Couette background flow. The growth rate of the large-scale
instability for the inhomogeneous turbulence with the "log-linear"
velocity shear is much larger than that for the Couette and
Poiseuille background flows. The characteristic spatial and time
scales for the instability are much larger than the characteristic
turbulent scales. This justifies separation of scales which is
required for the validity of the mean-field theory applied in the
present study.

The large-scale instability results in excitation of the turbulent
Tollmien-Schlichting waves. The  mechanism for the excitation of
these waves is different from that for the Tollmien-Schlichting
waves in laminar flows. In particular, the molecular viscosity plays
a crucial role in the excitation of the Tollmien-Schlichting waves
in laminar flows. Contrary, the turbulent Tollmien-Schlichting waves
are excited by a combined effect of the turbulent Reynolds
stress-induced generation of perturbations of the mean vorticity and
the background sheared motions. The energy of these waves is
supplied by the small-scale sheared turbulence, and the off-diagonal
terms in the turbulent viscosity tensor play a crucial role in the
excitation of the turbulent Tollmien-Schlichting waves.

Note that this study is principally different from the problems of
transition to turbulence whereby the stability of the laminar
Couette and Poiseuille flows are investigated (see, e.g.,
\cite{DR81,SH01,CJJ03,BOH88,REM03,ESH07}, and references therein).
Here we do not analyze a transition to turbulence. We study the
large-scale instability caused by an effect of the small-scale
anisotropic turbulence on the mean flow. This anisotropic turbulence
is produced by an interaction of equilibrium large-scale Couette or
Poiseuille flows with a small-scale isotropic background turbulence
produced by, e.g., a steering force. The anisotropic velocity
fluctuations are generated by tangling of the mean-velocity
gradients with the velocity fluctuations of the background
turbulence \cite{EKR03,EKRZ02}.

The "tangling" mechanism is an universal phenomenon that was
introduced in \cite{W57,BH59} for a passive scalar and in
\cite{G60,M61} for a passive vector (magnetic field). The Reynolds
stresses in a turbulent flow with a mean velocity shear is another
example of tangling anisotropic fluctuations \cite{L67}. For
instance, these velocity fluctuations are anisotropic in the
presence of shear and have a steeper spectrum $\propto k^{-7/3}$
than, e.g., a Kolmogorov background turbulence (see, e.g.,
\cite{L67,WC72,SV94,IY02,EKRZ02}). The anisotropic velocity
fluctuations determine the effective force and the Reynolds stresses
in Eq.~(\ref{B15}). This is the reason for the new terms $\propto
\beta_n \, l_0^2$ appearing in Eqs.~(\ref{AA1})-(\ref{A2}).

The obtained results in this study may be of relevance in different
turbulent astrophysical, geophysical and industrial flows.
Turbulence with a large-scale velocity shear is a universal feature
in astrophysics and geophysics. In particular, the analyzed effects
may be important, e.g., in accretion disks, extragalactic clusters,
merged protostellar and protogalactic clouds. Sheared motions
between interacting clouds can cause an excitation of the
large-scale instability which results in generation of the mean
vorticity and formation of large-scale vortical structures (see,
e.g., \cite{P80,ZN83,C93}). Dust particles can be trapped by the
vortical structures to enhance agglomeration of material and
formation of particle clusters \cite{BS95,BR98,EKR98,CH00,JAB04}.

The suggested mechanism can be used in the analysis of the flows
associated with Prandtl's turbulent secondary flows (see, e.g.,
\cite{P52,B87}). However, in this study we have investigated only
simple physical mechanisms to describe an initial (linear) stage of
the formation of vortical structures. The simple models considered
in this study can only mimic the flows associated with turbulent
secondary flows. Clearly, the comprehensive numerical simulations of
the nonlinear problem are required for quantitative description of
the turbulent secondary flows.

\begin{acknowledgments}
This research was supported in part by the Israel Science Foundation
governed by the Israeli Academy of Science, and by the Israeli
Universities Budget Planning Committee (VATAT).
\end{acknowledgments}


\begin{thebibliography}{}

\bibitem {L83} H.J. Lugt, {\em Vortex Flow in Nature and Technology}
(J. Wiley and Sons, New York, 1983), and references therein.

\bibitem {P87} J. Pedlosky, {\it Geophysical Fluid Dynamics}
(Springer, New York, 1987), and references therein.

\bibitem {C94} A.J. Chorin, {\it Vorticity and Turbulence}
(Springer, New York, 1994), and references therein.

\bibitem {GLM97}
A. Glasner, E. Livne and B. Meerson, Phys. Rev. Lett. {\bf 78}, 2112
(1997).

\bibitem {T98}
A. Tsinober, Eur. J. Mech. B/Fluids  {\bf 17}, 421 (1998).

\bibitem {RAO98}
C. Reyl, T. M. Antonsen and E. Ott, Physica D {\bf 111}, 202 (1998).

\bibitem {P52} L. Prandtl, {\it Essentials of Fluid Dynamics}
(Blackie, London, 1952).

\bibitem {T56} A.A. Townsend, {\it The Structure of Turbulent Shear Flow}
(Cambridge Univ. Press, Cambridge, 1956).

\bibitem {P70}
H.J. Perkins, J. Fluid Mech. {\bf 44}, 721 (1970).

\bibitem {B87}
P. Bradshaw, Ann. Rev. Fluid Mech. {\bf 19}, 53 (1987).

\bibitem {EKR03} T. Elperin, N. Kleeorin and I. Rogachevskii,
Phys. Rev. E {\bf 68}, 016311 (2003).

\bibitem {MST83}
S.S. Moiseev, R.Z. Sagdeev, A.V. Tur, G.A. Khomenko, and A.M.
Shukurov, Sov. Phys. Dokl. {\bf 28}, 925 (1983) [Dokl. Acad. Nauk
SSSR {\bf 273}, 549 (1983)].

\bibitem {KMT91}
G.A. Khomenko, S.S. Moiseev and A.V. Tur, J. Fluid Mech. {\bf 225},
355 (1991).

\bibitem {CMP94}
O.G. Chkhetiany, S.S. Moiseev, A.S. Petrosyan and R.Z. Sagdeev,
Physica Scripta {\bf 49}, 214 (1994).

\bibitem {DR81}
P. G. Drazin and  W. H. Reid, {\it Hydrodynamic Stability}
(Cambridge Univ. Press, Cambridge, 1981).

\bibitem {SH01}
P. J. Schmid and D. S. Henningson, {\it Stability and Transition in
Shear Flows} (Springer, Berlin, 2001).

\bibitem {CJJ03}
W. O. Criminale, T. L. Jackson and R. D. Joslin, {\it Theory and
Computation of Hydrodynamic Stability} (Cambridge Univ. Press,
Cambridge, 2003).

\bibitem {BOH88}
B. J. Bayly,  S. A. Orszag and Th. Herbert, Annu. Rev. Fluid Mech.
{\bf 20}, 359  (1988).

\bibitem {REM03}
D. Rempfer, Annu. Rev. Fluid Mech. {\bf 35}, 229  (2003).

\bibitem {ESH07}
B. Eckhardt, T. M. Schneider, B. Hof and J. Westerweel, Annu. Rev.
Fluid Mech.  {\bf 39}, 447 (2007).

\bibitem {O70}  S. A. Orszag, J. Fluid Mech. {\bf 41}, 363 (1970).

\bibitem {MY75} A. S. Monin and A. M. Yaglom, {\it Statistical Fluid
Mechanics}  (MIT Press, Cambridge, Massachusetts, 1975).

\bibitem {Mc90} W.D. McComb, {\it The Physics of Fluid Turbulence}
(Clarendon,  Oxford, 1990).

\bibitem {PFL76} A. Pouquet, U. Frisch, and J. Leorat, J. Fluid Mech.
{\bf 77}, 321 (1976).

\bibitem {KRR90}
N. Kleeorin, I. Rogachevskii, and A. Ruzmaikin, Zh. Eksp. Teor. Fiz.
{\bf 97}, 1555 (1990) [Sov. Phys. JETP {\bf 70}, 878 (1990)].

\bibitem {EKRZ02} T. Elperin, N. Kleeorin, I. Rogachevskii
and S.S. Zilitinkevich, Phys. Rev. E {\bf 66}, 066305 (2002).

\bibitem {BF02} E. G. Blackman and G. Field,  Phys. Rev. Lett.
{\bf 89}, 265007 (2002); Phys. Fluids {\bf 15}, L73 (2003).

\bibitem {FB02} G. Field and E. G. Blackman,  Astrophys. J.
{\bf 572}, 685 (2002).

\bibitem {BK04}
A. Brandenburg, P. K\"{a}pyl\"{a}, and A. Mohammed, Phys. Fluids
{\bf 16}, 1020 (2004).

\bibitem {BSM05}
A. Brandenburg and K. Subramanian, Phys. Rept. {\bf 417}, 1 (2005);
Astron. Astrophys. {\bf 439}, 835 (2005).

\bibitem {SSB07}
S. Sur, K. Subramanian and A. Brandenburg, Mon. Not. Roy. Astron.
Soc. {\bf 376}, 1238 (2007).

\bibitem {RUK06}
G. R\"{u}diger and L. L. Kitchatinov, Astron. Nachr. {\bf 327}, 298
(2006).

\bibitem {RKL06}
I. Rogachevskii, N. Kleeorin and E. Liverts, Geophys. Astrophys.
Fluid Dynam. {\bf 100}, 537 (2006); I. Rogachevskii and N. Kleeorin,
Phys. Rev. E {\bf 75}, 046305 (2007).

\bibitem {W57}
A.D. Wheelon, Phys. Rev. {\bf 105}, 1706 (1957).

\bibitem {BH59}
G.K. Batchelor, I.D. Howells and A.A. Townsend, J. Fluid Mech. {\bf
5}, 134 (1959).

\bibitem {G60}
G.S. Golitsyn, Doklady Acad. Nauk {\bf 132}, 315 (1960) [Soviet
Phys. Doklady {\bf 5}, 536 (1960)].

\bibitem {M61}
H.K. Moffatt, J. Fluid Mech. {\bf 11}, 625 (1961).

\bibitem {L67}
J.L. Lumley, Phys. Fluids, {\bf 10} 1405 (1967).

\bibitem {WC72}
J.C. Wyngaard and O.R. Cote, Q. J. R. Meteorol. Soc. {\bf 98}, 590
(1972).

\bibitem {SV94}
S.G. Saddoughi and S.V. Veeravalli, J. Fluid Mech. {\bf 268}, 333
(1994).

\bibitem {IY02}
T. Ishihara, K. Yoshida and Y. Kaneda, Phys. Rev. Lett. {\bf 88},
154501 (2002).

\bibitem {P80}
P. J. E. Peebles, {\it The Large Scale Structure of the Universe},
(Princeton Univ. Press, Princeton, 1980).

\bibitem {ZN83}
Ya. B. Zeldovich and  I. D. Novikov, {\it Relativistic
Astrophysics}, Vol. 2, {\it The Structure and Evolution of the
Universe}, Chicago Univ. Press, Chicago, 1983).

\bibitem {C93}
A. D. Chernin, Astron. Astrophys.  {\bf 267}, 315 (1993); Astrophys.
Space Sci. {\bf 186}, 159 (1991).

\bibitem {BS95}
P. Barge and J.Sommeria, Astron. Astrophys. {\bf 295}, L1  (1995).

\bibitem {BR98} L. S. Hodgson and A. Brandenburg,
Astron. Astrophys. {\bf 330}, 1169 (1998).

\bibitem {EKR98} T. Elperin, N. Kleeorin and I. Rogachevskii,
Phys. Rev. Lett. {\bf 81}, 2898 (1998).

\bibitem {CH00} P. H. Chavanis,
Astron. Astrophys. {\bf 356}, 1089 (2000).

\bibitem {JAB04}
A. Johansen, A. C. Andersen and A. Brandenburg, Astron. Astrophys.
{\bf 417}, 361 (2004).

\end{thebibliography}
\end{document}